\documentclass[%
 reprint,
 amsmath,amssymb,
 aps,
]{revtex4-1}

\usepackage{graphicx}
\usepackage{dcolumn}
\usepackage{bm}

\begin{document}

\preprint{APS}

\title{High-pressure phase relations in Zn$_2$SiO$_4$ system:\\A first-principles study}

\author{Masami Kanzaki}
 \email{mkanzaki@okayama-u.ac.jp}
\affiliation{%
 Institute for Planetary Materials, Okayama University,\\
827 Yamada, Misasa, Tottori, Japan 682-0193
}%

\date{\today}

\begin{abstract}
Recent experimental studies have shown that phases III and IV of Zn$_2$SiO$_4$ recovered from high-pressure experiments are retrograde phases. In order to clarify the phase relation of this system, first-principles density functional theory calculations of 11 Zn$_2$SiO$_4$ phases including phases III and IV were conducted. Phase III, having a “tetrahedral olivine” structure, exhibited an extraordinarily high compressibility, which is due to large volume reductions in vacant octahedral sites corresponding to $\it{M}$1 and $\it{M}$2 sites in olivine structure. Both phases III and IV have much higher enthalpies compared to those of phases II and V up to 10 GPa, and they are not stable high-pressure phases. Instead, Na$_2$CrO$_4$- and Ag$_2$CrO$_4$-structured phases have volumes and enthalpies next to phase V at around 9 GPa, and they are most likely candidate structures for high-pressure phases of III and IV, respectively. In both structures, a half of Zn is occupying a tetrahedral site, and the remaining half is occupying an octahedral site. Compared to those phases, olivine phase has slightly higher volume and enthalpy, being a less likely candidate. ``Phase transitions'' of phases II, III and IV observed during structural optimization under pressure are also reported. The transition of phase II was discussed in relation to similar transition known for Si$_3$N$_4$.
\end{abstract}

\maketitle


\section{Introduction}

Zn$^{2+}$ ion prefers tetrahedral sites over octahedral sites  \cite{Syono}. Consequently, zinc silicates exhibit quite different crystalline phases and phase relation compared to those of magnesium silicates at ambient pressure. Partition coefficient of Zn between olivine and a basaltic magma significantly deviates from those of other divalent cations with similar ionic radii, resulting Zn anomaly \cite{Matsui}. Thus, crystal chemical understanding of zinc silicates might provide better insight of chemical behavior of Zn in nature and in synthetic materials. 

Pressure-induced phase transition sequence of Zn$_2$SiO$_4$ determined by previous quench experiments up to 15 GPa is as follow: I$\rightarrow$II$\rightarrow$III$\rightarrow$IV$\rightarrow$V. Phase V then decomposes to ZnSiO$_3$ ilmenite phase plus ZnO (NaCl structure) at around 13 GPa \cite{Ito}. Synthesis of phase VI (structure unknown) was reported at around 12 GPa \cite{Doroshev}, but the phase has not been confirmed by other studies.

Phase I (mineral name willemite) has a phenacite structure \cite{Klaska}, whereas phase II adopts a compact and rare structure \cite{Marumo}. In both structures, Zn and Si occupy tetrahedral sites. Crystal structures of phases III and IV are reported recently \cite{Liu}. Phase III adapts a “tetrahedral olivine” structure \cite{Baur} in which Zn ions occupying vacant tetrahedral sites of olivine structure, while leaving $\it{M}$1 and $\it{M}$2 sites of olivine structure vacant. Phase IV consists of tetrahedrally coordinated Zn and Si, and it features unique edge-shared Zn$_2$O$_6$ dimers. This study also reveals that these two phases have volumes lower than that of lower pressure phase (i.e., phase II), and the volume of phase III is very close to that of the ambient pressure phase (phase I). Therefore, these phases are apparently not high-pressure phases, and are likely transformed from yet unknown high-pressure phases during the decompression process. This sort of the retrograde phase transformation of high-pressure phase during decompression to a metastable phase are known for other compounds \cite{Yusa}. Phase V was identified to have modified spinel structure \cite{Syono}, and its crystal structure was refined recently \cite{Kanzaki2018a}. This is the only experimentally known Zn$_2$SiO$_4$ phase in which all Zn ions occupy octahedral sites in the structure.

Recent in-situ high-pressure Raman spectroscopic study of phases III and IV at room temperature revealed that both phases exhibited the pressure-induced phase transitions at 5.5 and 2.5 GPa during compression, respectively \cite{Kanzaki2018c}. These transformed phases are abbreviated as III-HP and IV-HP hereafter to avoid confusion. This study certified that phases III and IV are retrograde phases from the supposed III-HP and IV-HP, respectively. However, the crystal structures of phases III-HP and IV-HP are not known yet. In order to understand stabilities of these phases, and to find structure candidates for these structure-unknown phases, first-principles density functional theory (DFT) calculations were conducted in present study. The DFT study of seven phases in this system has been reported before \cite{Karazhanov}. Their calculations included structures of $\beta$-Ca$_2$SiO$_4$ (mineral name larnite) and olivine which they regarded as model structures for phases III and IV, respectively. However, as shown by Liu et al. \cite{Liu}, crystal structures of phases III and IV are different from those model structures. Therefore, new thorough calculations including these newly established structures and other possible candidate structures are needed. Nalbandyan and Novikova \cite{Nalbandyan} reviewed structural chemistry of A$_2$MX$_4$ compounds in terms of packing densities, and their result can be used to estimate high-pressure candidate phases for Zn$_2$SiO$_4$ compounds. In the present study, structural optimizations of 11 selected phases were conducted up to 25 GPa using the first-principles DFT method.

\section{Calculation method}
The first-principles DFT calculations were conducted using the QUANTUM ESPRESSO (ver. 6.2) package \cite{qespresso2}. Projector augmented-wave (PAW) method was used with following potentials: Zn.pbesol-dn-kjpaw$\_$psl.0.2.2.UPF, Si.pbesol-n-kjpaw$\_$psl.0.1.UPF, O.pbesol-n-kjpaw$\_$psl.0.1.UPF from PSlibary \cite{DalCorso}. Following 21 phases were considered at the very beginning: I, II, III, IV, V, olivine, spinel, III-$\it{Pbam}$, Na$_2$CrO$_4$, Ag$_2$CrO$_4$, Ca$_2$RuO$_4$, thenardite (Na$_2$SO$_4$), $\beta$-Li$_2$SO$_4$, larnite, K$_2$SO$_4$, Na$_2$MnO$_4$, K$_2$MnO$_4$, K$_2$MoO$_4$, Tl$_2$CrO$_4$, NaGdSiO$_4$ and Zn$_2$SiO$_4$-$\it{m}$ (a phase reported in Karazhanov et al. \cite{Karazhanov}). However, last 10 phases have enthalpies substantially higher than other phases at 0 and 9 GPa. Therefore, only first 11 phases are considered further. Ilmenite phase of ZnSiO$_3$ was also calculated to compare with experimental compression data. 

The kinetic energy cutoff for wave function, the kinetic energy cutoff for charge density, and the scf convergence threshold were set 80, 400 and 10$^{-14}$ Ry, respectively. The Brillouin zones were sampled with the Monkhorst-Pack scheme. Used meshes were 4$\times$4$\times$4 for phase II, spinel and ilmenite, 4$\times$2$\times$8 for phases III and III-$\it{Pbam}$ (a high-pressure phase transformed from phase III, see below), 2$\times$4$\times$4 for olivine and Ag$_2$CrO$_4$,  2$\times$2$\times$2 for phase I, 2$\times$2$\times$4 for phase IV, 4$\times$2$\times$2 for phase V, 4$\times$2$\times$4 for Na$_2$CrO$_4$ and 4$\times$4$\times$2 for Ca$_2$RuO$_4$. At each pressure, atomic positions and cell parameters are optimized using BEFG algorithm \cite{qespresso2}. Zero-point vibrational energy was not included in the enthalpy calculation.

Initial crystal structural parameters were taken from experimental ones for I, II, III, IV and V \cite{Klaska,Marumo,Liu,Kanzaki2018a}. For olivine, spinel and ilmenite phases, structures of Mg-silicate polymorphs were used \cite{Fujino,Sasaki,Horiuchi}, and Mg positions are replaced with Zn. For   Na$_2$CrO$_4$ and Ag$_2$CrO$_4$ phases, isostructural Na$_2$SO$_4$ polymorphs \cite{Rasmussen} were used, and Na and S are replaced with Zn and Si, respectively. For Ca$_2$RuO$_4$ structure, the structural parameters were taken from Karazhanov et al. \cite{Karazhanov}. The equations of states from the calculated volumes were fitted to the third-order Birch-Murnaghan equations using the non-linear fitting function (nls) of R \cite{R}.

\section{Results and discussion}
\subsubsection{Structural optimizations}
The compression curves of 11 phases up to 25 GPa are shown in Figure 1, and the parameters for the equation of state are listed in Table 1. Because of ``transition'' during structural optimization, some curves are terminated below 25 GPa. The optimized structures of these phases at 1 bar and 0 K are given in Table 2. Comparison with available experimental cell volumes at ambient pressure for phases I, II, III, IV and V showed that the calculated volumes are slightly smaller than observed ones, mostly within 0.6\% (Table 1). We noted that the pbesol-type pseudopotentials which use general gradient approximation reproduce the cell volume well. Karazhanov et al. \cite{Karazhanov} conducted the DFT calculations using local density approximation (LDA). Their calculated volumes at ambient pressure for phases I, II and V are a few \% smaller than those of experiments and are typical for the LDA calculations \cite{Kroll}. 

\begin{figure}[bt]
 \begin{center}
  \includegraphics[clip,width=8.5cm]{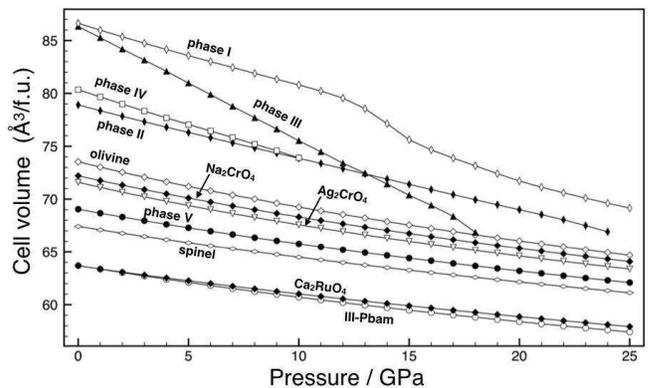}
  \caption{Compression curves of eleven Zn$_2$SiO$_4$ phases}
 \end{center}
\end{figure}

\begin{table}[hbtp]
\caption{Equations of states of Zn$_2$SiO$_4$ phases}
\begin{tabular}{clllll}
\hline
phase & V$_0$\footnotemark[1](\AA$^{3}$) & V(obs)  & K$_0$(GPa) & K$^{\prime}$ & P range \\
\hline
I & 86.607(12) & 86.92 & 137.0(11) & 1.6(2) & 0--11 \\
II & 78.894(12) & 79.34 & 149.4(6) & 0.63(4) & 0--23 \\
III & 86.58(8) & 87.03 & 70.7(11) & 1.10(8) & 0--17 \\
IV & 80.33(2) & 80.85 & 119.5(14) & 0.6(2) & 0--10 \\
V & 69.037(2) & 69.30 & 183.1(2) & 4.76(2) & 0--25 \\
olivine & 73.515(4) & 74.3\footnotemark[2] & 145.9(3) & 4.56(3) & 0--25 \\
spinel & 67.4145(4) & NA & 204.21(5) & 4.561(4) & 0--25 \\
Na$_{2}$CrO$_{4}$ & 72.198(2) & NA & 158.9(2) & 4.63(2) & 0--25 \\
Ag$_{2}$CrO$_{4}$ & 71.577(4) & NA & 149.5(3) & 5.23(3) & 0--25 \\
Ca$_{2}$RuO$_{4}$ & 63.666(2) & NA & 213.7(3) & 4.52(3) & 0--25 \\
III-$\it{Pbam}$ & 63.714(4) & NA & 181.4(5) & 5.37(5) & 0--25 \\
ilmenite & 44.9017(14) & 44.73 & 204.3(3) & 4.97(3) & 0--25 \\
\hline
\footnotetext[1]{Volume per formula}
\footnotetext[2]{Estimated from Mg-Zn olivine solid solution \cite{Syono}}
\end{tabular}
\end{table}

\begin{table*}[bt]
\caption{Optimized crystal structures of Zn$_2$SiO$_4$ phases at 1 bar and 0 K}
\begin{tabular}{|lccc|lccc|lccc|}
\hline
\multicolumn{4}{|c|}{willemite} & \multicolumn{4}{c|}{phase II} & \multicolumn{4}{c|}{phase III}  \\
\hline
\multicolumn{4}{|c|}{$\it{R}$$\bar
{3}$, $\it{a}$ = 8.6197 \AA, } & \multicolumn{4}{c|}{ $\it{I}$$\bar{4}$2$\it{d}$, $\it{a}$ = 7.0124,} & \multicolumn{4}{c|}{$\it{Pnma}$, $\it{a}$ = 10.2804, }  \\
\multicolumn{4}{|c|}{$\alpha$ = 107.896$^\circ$} & \multicolumn{4}{c|}{$\it{c}$ = 6.4193 \AA} & \multicolumn{4}{c|}{$\it{b}$ = 6.6360, $\it{c}$ = 5.0609 \AA}  \\
\hline
site & x & y & z & site & x & y & z & site & x & y & z \\
\hline
Zn1(6f) & 0.3986 & 0.6255 & 0.2231 & Zn(8d) & 0.1565 & 1/4 & 1/8 & Zn(8d) & 0.1562 & 0.9996 & 0.8248 \\
Zn2(6f) & 0.0591 & 0.2961 & 0.8897 & Si(4b) & 0 & 0 & 1/2 & Si(4c) & 0.0923 & 1/4 & 0.3231 \\
Si(6f) & 0.7339 & 0.9611 & 0.5539 & O(16e) & 0.3060 & 0.4845 & 0.1426 & O1(4c) & 0.1120 & 1/4 & 0.6446 \\
O1(6f) & 0.8605 & 0.8543 & 0.5346 &  &  &  &  & O2(4c) & 0.4392 & 1/4 & 0.2631 \\
O2(6f) & 0.7483 & 0.0736 & 0.4325 &  &  &  &  & O3(8d) & 0.1632 & 0.0480 & 0.2051 \\
O3(6f) & 0.8114 & 0.1017 & 0.7686 &  &  &  &  & & & & \\
O4(6f) & 0.5250 & 0.8090 & 0.4743 &  &  &  &  & & & & \\
\hline
\hline
\multicolumn{4}{|c|}{phase IV} & \multicolumn{4}{c|}{olivine phase} & \multicolumn{4}{c|}{phase V}  \\
\hline
\multicolumn{4}{|c|}{$\it{Pbca}$, $\it{a}$ = 10.8823, } & \multicolumn{4}{c|}{$\it{Pnma}$, $\it{a}$ = 10.2520,} & \multicolumn{4}{c|}{$\it{Imma}$, $\it{a}$ = 5.7358, }  \\
\multicolumn{4}{|c|}{$\it{b}$ = 9.7025, $\it{c}$ = 6.0886 \AA,} & \multicolumn{4}{c|}{$\it{b}$ = 6.0014, $\it{c}$ = 4.7805 \AA} & \multicolumn{4}{c|}{$\it{b}$ = 11.4936, $\it{c}$ = 8.3783 \AA}  \\
\hline
site & x & y & z & site & x & y & z & site & x & y & z \\
\hline
Zn1(8c) & 0.5618 & 0.5819 & 0.3458 & Zn1(4a) & 0 & 0 & 0 & Zn1(4a) & 0 & 0 & 0 \\
Zn2(8c) & 0.6606 & 0.3165 & 0.0564 & Zn2(4c) & 0.2797 & 1/4 & 0.9846 & Zn2(4e) & 0 & 1/4 & 0.9694 \\
Si(8c) & 0.3760 & 0.1200 & 0.6366 & Si(4c) & 0.0958 & 1/4 & 0.4268 & Zn3(8g) & 1/4 & 0.1259 & 1/4 \\
O1(8c) & 0.2515 & 0.2162 & 0.6686 & O1(4c) & 0.0933 & 1/4 & 0.7673 & Si(8h) & 0 & 0.1204 & 0.6172 \\
O2(8c) & 0.4008 & 0.0232 & 0.8517 & O2(4c) & 0.4472 & 1/4 & 0.2176 & O1(4e) & 0 & 1/4 & 0.2141 \\
O3(8c) & 0.4882 & 0.2302 & 0.5944 & O3(8d) & 0.1647 & 0.0316 & 0.2791 & O2(4e) & 0 & 1/4 & 0.7161 \\
O4(8c) & 0.3642 & 0.0185 & 0.4225 &  &  &  &  & O3(8h) & 0 & 0.9887 & 0.2563 \\
 &  &  &  &  &  &  &  & O4(16j) & 0.2625 & 0.1230 & 0.9922 \\
\hline
\hline
\multicolumn{4}{|c|}{spinel phase} & \multicolumn{4}{c|}{Na$_{2}$CrO$_{4}$ phase} & \multicolumn{4}{c|}{Ag$_{2}$CrO$_{4}$ phase}  \\
\hline
\multicolumn{4}{|c|}{$\it{Fd}$$\bar{3}$$\it{m}$, $\it{a}$ = 8.1397 \AA } & \multicolumn{4}{c|}{$\it{Cmcm}$, $\it{a}$ = 5.5743,} & \multicolumn{4}{c|}{$\it{Pnma}$, $\it{a}$ = 9.3999,}  \\
\multicolumn{4}{|c|}{ } & \multicolumn{4}{c|}{$\it{b}$ = 8.5267, $\it{c}$ = 6.0766 \AA} & \multicolumn{4}{c|}{$\it{b}$ = 6.1644, $\it{c}$ = 4.9514 \AA}  \\
\hline
site & x & y & z & site & x & y & z & site & x & y & z \\
\hline
Zn(16d) & 5/8 & 5/8 & 5/8 & Zn1(4b) & 0 & 1/2 & 0 & Zn1(4b) & 1/2 & 0 & 0 \\
Si(8a) & 0 & 0 & 0 & Zn2(4c) & 0 & 0.1660 & 1/4 & Zn2(4c) & 0.1659 & 1/4 & 0.9871 \\
O(32e) & 0.3681 & 0.3681 & 0.3681 & Si(4c) & 0 & 0.8513 & 1/4 & Si(4c) & 0.8216 & 1/4 & 0.0252 \\
 &  &  &  & O1(8g) & 0.2691 & 0.4754 & 1/4 & O1(4c) & 0.6488 & 1/4 & 0.1107 \\
 &  &  &  & O2(8f) & 0 & 0.2611 & 0.5336 & O2(4c) & 0.8582 & 1/4 & 0.7000 \\
 &  &  &  & & & & & O3(8d) & 0.8869 & 0.0323 & 0.1670 \\
\hline
\hline
\multicolumn{4}{|c|}{III-$\it{Pbam}$ phase} & \multicolumn{4}{c|}{Ca$_{2}$RuO$_{4}$ phase} & \multicolumn{4}{c|}{}  \\
\hline
\multicolumn{4}{|c|}{$\it{Pbam}$, $\it{a}$ = 4.9128 \AA } & \multicolumn{4}{c|}{$\it{Pbca}$, $\it{a}$ = 4.9236,} & \multicolumn{4}{c|}{}  \\
\multicolumn{4}{|c|}{$\it{b}$ = 9.3540, $\it{c}$ = 2.7730 \AA} & \multicolumn{4}{c|}{$\it{b}$ = 5.0143, $\it{c}$ = 10.3174 \AA} & \multicolumn{4}{c|}{}  \\
\hline
site & x & y & z & site & x & y & z & & & & \\
\hline
Zn(4g) & 0.9688 & 0.1833 & 0 & Zn(8c) & 0.9989 & 0.05376 & 0.3364 &  &  &  &  \\
Si(2d) & 1/2 & 0 & 1/2 & Si(4a) & 0 & 0 & 0 &  &  &  & \\
O1(4g) & 0.2856 & 0.0559	 & 0 & O1(8c) & 0.1972 & 0.2925 & 0.0479 &  &  &  & \\
O2(4h) & 0.6668 & 0.1715 & 1/2 & O2(8c) & 0.8874 & 0.9532 & 0.1598 &  &  &  & \\
\hline
\end{tabular}
\end{table*}

There is no experimental compression study for any of Zn$_2$SiO$_4$ phases so far, but the compression data is available for ZnSiO$_3$ ilmenite phase up to 12 GPa \cite{Sato}. Therefore, the compression curve of ilmenite phase was also calculated to assess reproducibility of our calculation. The reported bulk modulus (216 GPa with fixed K$^{\prime}$=4) can compare well with our calculated one (204.2 GPa with K$^{\prime}$=4.97), whereas Karazhanov et al. \cite{Karazhanov} reported much smaller bulk modulus (177.89 GPa with K$^{\prime}$= 5.5). Sato et al. \cite{Sato} stated that their pressure scale based on LiF lattice parameter is likely overestimated pressure by 5\% compared to more popular NaCl scale \cite{Decker}. If this is the case, their bulk modulus is recalculated as 210 GPa, resulting further agreement with our calculated value. Therefore, similar accuracy in calculated compressibility would be expected for the Zn$_2$SiO$_4$ system too.

The compression curves in Figure 1 demonstrate a general trend that a phase with larger volume is more compressible. Among them, phases I and III revealed anomalous compressional behavior. The compression curve of phase I has a kink at around 14 GPa. Close inspection of the structure reveals that there is no change in the space group ($\it{R}$$\bar{3}$) up to 25 GPa. Figure 2 shows lattice parameters, tetrahedral volumes and selected inter-tetrahedral angles of phase I with pressure. As shown in Figure 2d, the inter-tetrahedral angles change more effectively above 15 GPa, contributing reduction of void space made by six-membered rings. Therefore, this kink is likely due to a change in dominant compressional mechanism.

Phase III exhibits an unusual low bulk modulus, and is less than a half compared to those of other phases, except phase IV (see Table 1). At ambient pressure, phase III has a volume comparable to that of phase I. However, it surpasses phase II at 14 GPa and almost reaches to that of olivine phase at 18 GPa, where a ``transition'' happens. Liu et al. \cite{Liu} revealed that phase III has tetrahedral olivine structure \cite{Baur}. ``Normal'' (Mg,Fe)$_2$SiO$_4$ olivine structure is based on hexagonal closed packing of oxygens, and (Mg,Fe) ions occupy two octahedral sites ($\it{M}$1 and $\it{M}$2), and Si ion occupies a tetrahedral site. Tetrahedral olivine is named by Baur \cite{Baur}, and it can be derived from normal olivine structure by relocating cations in $\it{M}$1 and $\it{M}$2 sites into one of vacant tetrahedral sites (see Figure 4a for crystal structure of phase III). 

Therefore, it is interesting to compare structural changes of phase III with olivine phase. Figure 3 compares the lattice parameters of both structures with pressure. The $\it{a}$- and $\it{c}$-axis of phase III show higher linear compressibilities than those of olivine phase. Figure 3 shows polyhedral volumes of the cation sites of two phases with pressure. It also includes two vacant octahedral sites of phase III corresponding to $\it{M}$1 and $\it{M}$2 sites in olivine structure, and a vacant tetrahedral site in normal olivine, which is occupied by Zn in phase III. Significant reduction in volume for the vacant octahedral sites in phase III is apparent in Figure 3e, while nearly similar changes in the tetrahedral volumes of both occupied and vacant sites in Figure 3d. At 18 GPa, volumes of two vacant octahedral sites in phase III approach to those of $\it{M}$1 and $\it{M}$2 sites in olivine phase, a trend parallel with the volume change (Figure 1). These results suggest that the high compressibilities of the vacant octahedral sites are responsible for the anomalous bulk compressibility of phase III. 

Phase III was observed in the experimental product recovered from 8 to 10 GPa \cite{Syono}. Even though phase III has the high compressibility, the volume of phase III is still higher than that of phase II at those pressures, confirming that phase III is not high-pressure phase. Phase IV showed the compressibility higher than those of other phases, except phase III, and a transition during optimization was found at 11 GPa. Phase IV was observed in experimental products recovered from 9 to 11 GPa \cite{Syono, Ito}. At 10 GPa, the volume of phase IV is nearly same to that of phase II, but much larger compared to that of phase V. 

Since we are looking for the candidate structures of phases III-HP and IV-HP, the curves locating between those of phase II and phase V in Figure 1 are of special interest. They are olivine, Na$_2$CrO$_4$ and Ag$_2$CrO$_4$ phases. These phases are also listed as possible high-pressure phases of A$_2$BX$_4$ system (with tetrahedral B ion) \cite{Nalbandyan}. There are other possible candidate phases (such as thenardite) listed in Nalbandyan and Novikova \cite{Nalbandyan}. Those phases generally have large cation sites for A cation, and have much higher enthalpies. Na$_2$CrO$_4$ and Ag$_2$CrO$_4$ structures are not adopted in other M$_2$SiO$_4$ compounds (with M = Mg and the transition metal cations) and are not considered as the candidate high-pressure phases of Zn$_2$SiO$_4$ before.

\begin{figure}[bt]
 \begin{center}
  \includegraphics[clip,width=8.5cm]{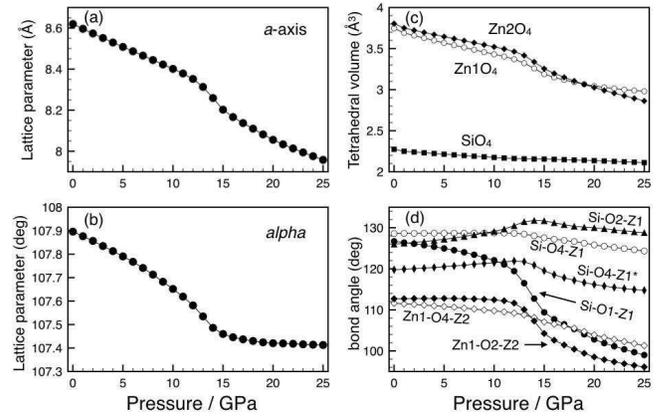}
  \caption{Cell parameters and structural changes with pressure for phase I. (a) and (b) cell parameters $\it{a}$ and $\it{\alpha}$, respectively; (c) tetrahedral volumes, (d) inter-tetrahedral angles in the 6-membered ring parallel to 111, except Si-O4-Z1*.}
 \end{center}
 
\end{figure}
\begin{figure}[bt]
 \begin{center}
  \includegraphics[clip,width=8.5cm]{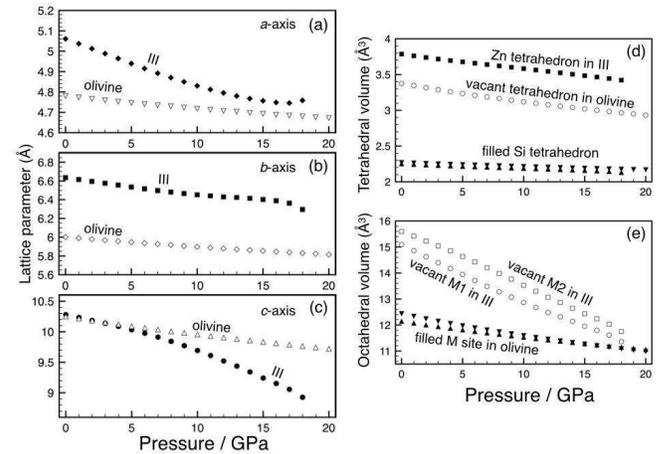}
  \caption{Cell parameters and polyhedral volumes of phase III and olivine with pressure. (a), (b) and (c): cell parameters $\it{a}$, $\it{b}$ and $\it{c}$, respectively. Note that for comparison, $\it{Pnma}$ space group with standard setting is used for both phases (see Table 2). (d) tetrahedral volumes: open circles represent volumes of the vacant tetrahedral site in olivine phase, and  solid squares represent volumes of the filled tetrahedral site in phase III. Filled up-pointing and down-pointed triangles show volumes of Si filled tetrahedral sites in phase III and olivine phase, respectively. e) octahedral volumes; open symbols show volumes of vacant M sites in phase III, and filled up-pointing and down-pointed triangles show volumes of $\it{M}$1 and $\it{M}$2 octahedra in olivine, respectively.}
 \end{center}
\end{figure}

\subsubsection{``Phase transitions'' observed during structure optimizations}
For phases II, III and IV, we observed pressure-induced “phase transition” during the structural optimization. Since the structural optimization is not exactly following evolution of dynamics of atoms with time, such as molecular dynamics simulations, these transitions should be treated with caution. Nevertheless, it would suggest prospective phase transitions. Therefore, these observed ``phase transitions'' are briefly described below.

\begin{figure*}[bt]
 \begin{center}
  \includegraphics[clip,width=16cm]{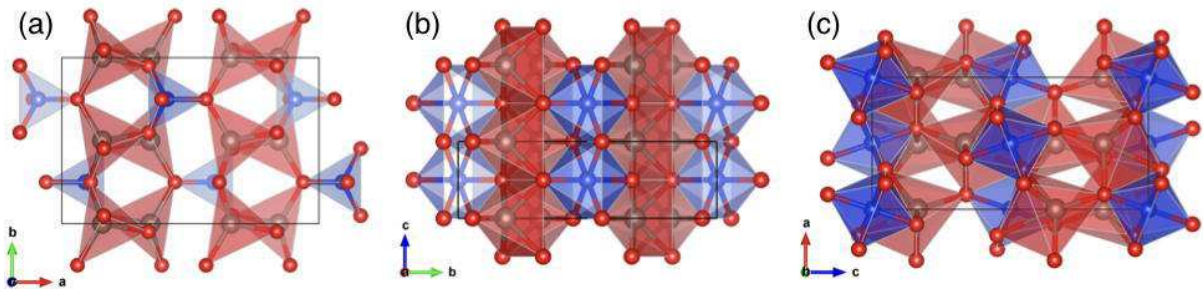}
  \caption{Crystal structures of Zn$_2$SiO$_4$ phases optimized by the DFT calculations at 1 bar and 0 K. Blue polyhedra represent Si sites, and red polyhedra represent Zn sites. (a) phase III, (b) phase III-$\it{Pbam}$, (c) Ca$_2$RuO$_4$ structure. Crystal structures are drawn by Vesta \cite{Vesta}.}
 \end{center}
\end{figure*}

Phase II was stable up to 24 GPa, but at 25 GPa, its volume dropped significantly during the optimization, and the obtained volume was comparable to that of spinel phase. Structure examination reveals that the resultant structure is indeed spinel structure. Spinel phase did not revert to phase II during the optimization even at 1 bar. This transition is discussed separately in the following section.

For phase III, a phase transition was observed at 19 GPa. Close examination of the resultant structure revealed that $\it{b}$-axis becomes half of original cell, and space group changed from $\it{Pnma}$ to $\it{Pcma}$ (= $\it{Pbam}$ in standard setting). Crystal structure of the phase (designated hereafter as phase III-$\it{Pbam}$) is shown in Figure 4 along with that of phase III, and atomic positions are given in Table 2. The structure consists of an octahedral Si site and a penta-coordinated Zn site. Penta-coordinated ZnO$_5$ forms a prism-like polyhedron. For penta-coordinated Zn in the phase, bond distance for sixth nearest oxygen is longer than 2.3 \AA, and it was not regarded as octahedral coordination. To our knowledge, no analogue phase with this structure is found in other high-pressure silicates, but we noted some structural similarity with Ca$_2$IrO$_4$ \cite{Ca2IrO4}. These polyhedral units are common with Ca$_2$RuO$_4$ structure too. However, SiO$_6$ octahedra are linked by edge-sharing, and form one-dimensional chains in III-$\it{Pbam}$, whereas SiO$_6$ octahedra are linked by sharing corners, and form a layer in Ca$_2$RuO$_4$. These two phases show very similar compressional behaviors (Figure 1). Kanzaki \cite{Kanzaki2018c} observed a phase transition of phase III to III-HP at 5.5 GPa during compression. So natural question is that phase III-$\it{Pbam}$ corresponds to phase III-HP? We are negative to this question. Experimentally observed transition pressure (5.5 GPa) is too low to realize octahedral Si in the structure. Therefore, phase III-$\it{Pbam}$ will not be a stable phase in this system. Phase III-$\it{Pbam}$) did not revert to phase III during the optimization even at 1 bar. 

For phase IV, a phase transition was observed at 10 GPa during the optimization. The resultant structure was found to be isostructural to Na$_2$CrO$_4$ structure with $\it{Cmcm}$ space group, and the structure was already included in our DFT calculations. Hereafter, the phase is designated as Na$_2$CrO$_4$ phase. As we will discuss later, we are not sure this phase corresponds to phase IV-HP or not. One of Na$_2$SO$_4$ phases (Na$_2$SO$_4$ III) also takes this structure \cite{Rasmussen}. Crystal structures of phase IV and Na$_2$CrO$_4$ are shown in Figure 5. It is noted that slight displacements of Zn ions in phase IV would result the structure of Na$_2$CrO$_4$ as shown by arrows in Figure 5a. In this structure, Zn ions occupy a tetrahedral and an octahedral sites, and Si ions occupy single tetrahedral site. Unique feature of this structure is edge-shared ZnO$_4$ and SiO$_4$ tetrahedra. Optimized structural parameters of this phase are given in Table 2. Na$_2$CrO$_4$ phase did not revert to phase IV during the optimization even at 1 bar.

Despite a half of Zn ions are still in the tetrahedral site, Na$_2$CrO$_4$ phase is denser than olivine phase in Zn$_2$SiO$_4$ system as shown in Figure 1, suggesting the phase could be a high-pressure phase of olivine phase in certain A$_2$BX$_4$ compounds. Such transition is actually known for LiFePO$_4$. LiFePO$_4$ adapts olivine structure at ambient pressure. Na$_2$CrO$_4$ phase of LiFePO$_4$ was synthesized at 6.5 GPa and 1173 K \cite{LiFePO4}. It is interesting to note that coordination number of Li is reduced from 6 to 4 by this transition, contrary to the well-known trend of pressure-induced coordination number increase. This structure was also observed as a high-pressure phase of Li$_2$SO$_4$ at 7 GPa \cite{Parfitt}. However, no isostructural phase is known for silicates so far.

\begin{figure*}[bt]
 \begin{center}
  \includegraphics[clip,width=16cm]{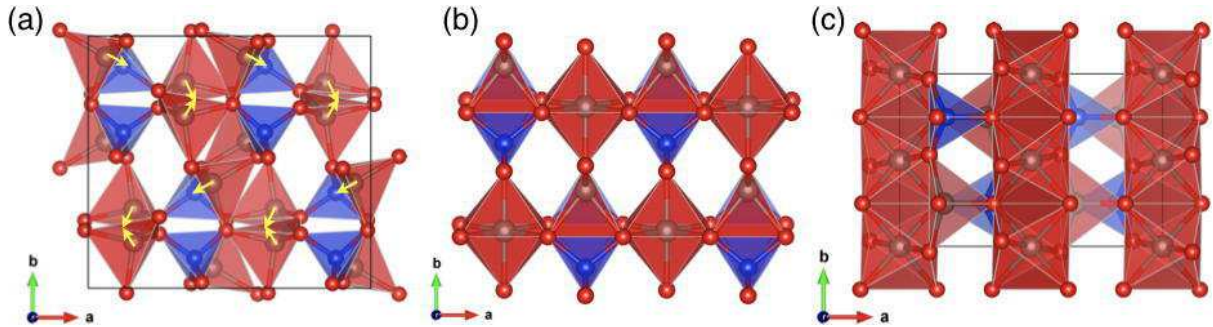}
  \caption{Crystal structures of Zn$_2$SiO$_4$ phases optimized by the DFT method at 1bar and 0 K. Blue polyhedra show Si sites, and red polyhedra represent Zn sites. (a) phase IV structure (allows indicate Zn movements during the transition), (b) Na$_2$CrO$_4$ structure (a candidate structure for high-pressure form of phase III), (c) Ag$_2$CrO$_4$ structure (a candidate structure for high-pressure form of phase IV). Structures are drawn by Vesta \cite{Vesta}.}
 \end{center}
\end{figure*}

\subsubsection{Stability of phases and possible high-pressure candidates}
Stability of phases can be evaluated by the enthalpies of 11 phases (at 0 K) as shown in Figure 6. The figure shows the enthalpies relative to that of phase II (always zero). Because phase II transformed to spinel phase at 25 GPa, the relative enthalpies are shown up to 24 GPa. At ambient pressure, phase I (willemite) has the lowest enthalpy, consistent with the experimental observation that phase I is the stable ambient pressure phase. Although phase III is highly compressible, it never becomes lowest enthalpy at any pressures. Situation is similar for phase IV. Thus, present calculations confirmed again that phases III and IV are metastable phases. At ambient pressure, phases III has second lowest enthalpy after phase I, and IV has fourth lowest enthalpy. This would explain why those phases appear as the retrograde phases during decompression.

\begin{figure*}[bt]
 \begin{center}
  \includegraphics[clip,width=16cm]{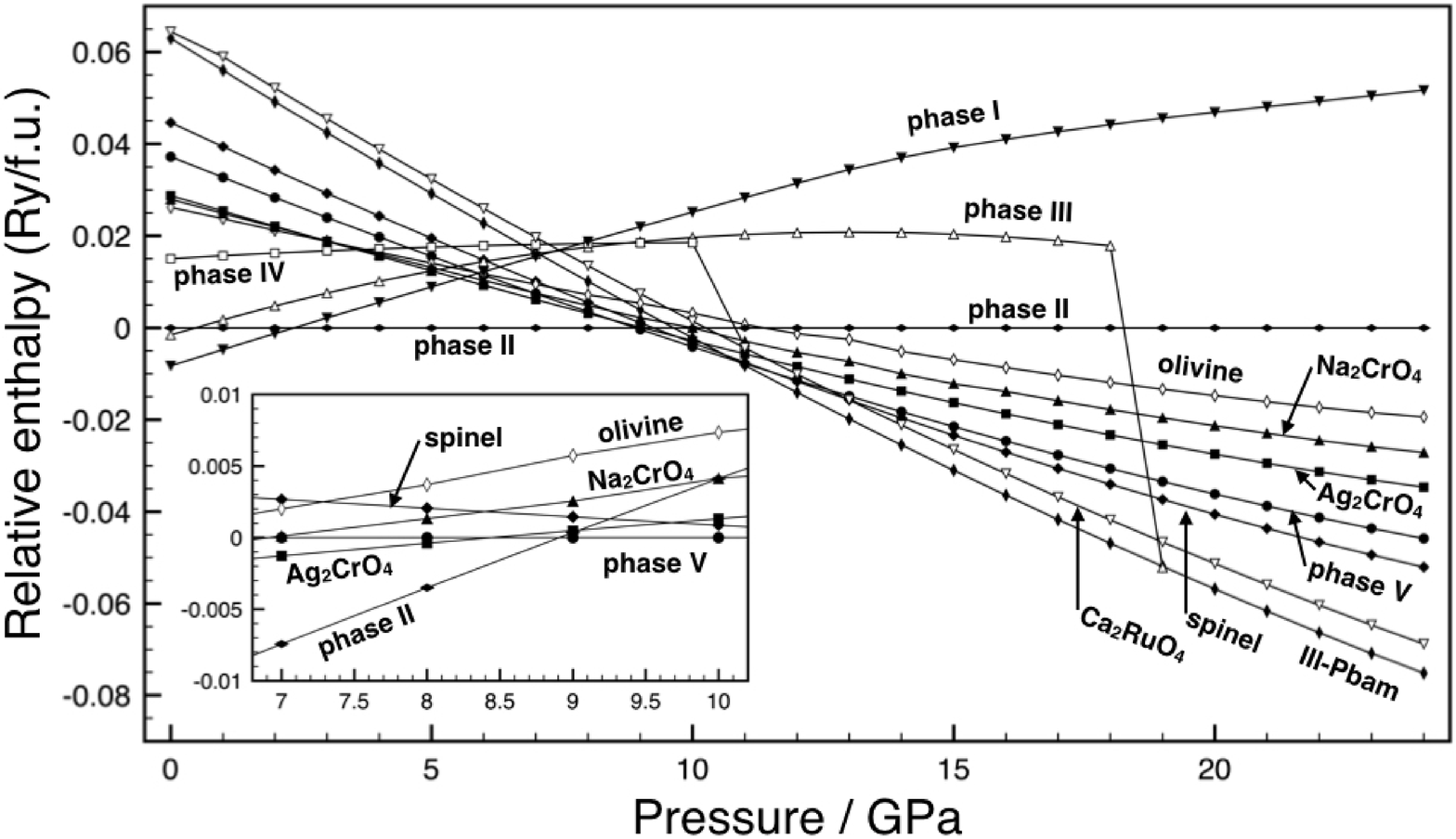}
  \caption{The enthalpies of 11 Zn$_2$SiO$_4$ phases relative to that of phase II with pressure at 0 K. Inset shows blow-up of 7--10 GPa region with the enthalpies relative to phase V.}
 \end{center}
\end{figure*}

At 3 GPa, phase II becomes the lowest enthalpy phase. Experimental phase boundary between phases I and II is about 3.5 GPa at 1000 $^\circ$C \cite{Syono}, and is consistent with present calculation. At 9 GPa, the enthalpy of phase V becomes lowest (see inset of Figure 6). Although lower pressure stability field of phase V is not well constrained experimentally, phase V has been synthesized at pressure as low as 11 GPa \cite{Ito}. This again is roughly consistent with the present calculation. Above 12 GPa, spinel, Ca$_2$RuO$_4$, and III-$\it{Pbam}$ phases have the enthalpies lower than that of phase V. However, phase V is experimentally known to decompose to ZnSiO$_3$ ilmenite plus ZnO at about 13 GPa (e.g., \cite{Ito}), so spinel phase and other dense phases (III-$\it{Pbam}$ and Ca$_2$RuO$_4$) cannot be appeared as a stable phase in this system.

Although the DFT calculation reproduced the observed phase relation of I, II and V well, there are missing phases between phases II and V: the high-pressure phases of III and IV (i.e., III-HP and IV-HP). As noted before, phases III and IV are ``experimentally'' observed at a pressure range between 8 to 11 GPa in the recovered samples \cite{Syono}. In-situ Raman spectroscopic study \cite{Kanzaki2018c} has demonstrated that phases III and IV transformed to the structure-unknown high-pressure phases (III-HP and IV-HP) at 5.5 and 2.5 GPa during compression, respectively. Therefore, there must be two stable phases corresponding those phases between phases II and V. However, no lowest enthalpy phases other than phase V were detected at pressure corresponding to synthesis of phases III and IV (8 to 11 GPa). It should be noted that our calculations are based on the enthalpies at 0 K without considering vibrational entropy contribution. Therefore, we need to consider not only lowest enthalpy phase, but also those phases which have enthalpies next to lowest one, considering the entropy contribution might stabilize those phases. At 7--10 GPa, the enthalpies of selected phases are closely compared (inset of Figure 6). This figure shows the enthalpies relative to phase V, and phase V becomes stable directly from phase II. However, the figure also revealed that the enthalpy of Ag$_2$CrO$_4$ phase is close to the enthalpy of phase V, and Na$_2$CrO$_4$ phase has slightly higher enthalpy, next to Ag$_2$CrO$_4$ phase. Olivine phase has the enthalpy always higher than those of Na$_2$CrO$_4$ and Ag$_2$CrO$_4$ phases. The compression curves (Figure 1) showed same order: olivine $>$ Na$_2$CrO$_4$ $>$ Ag$_2$CrO$_4$ $>$ phase V. Considering these, we  suggest following assignments: phase III-HP would have Na$_2$CrO$_4$ structure, whereas phase IV-HP would take Ag$_2$CrO$_4$ structure. Crystal structures of phase IV and Ag$_2$CrO$_4$ phase are shown in Figure 5.

Although this suggestion seems reasonable, there is one inconsistency still remains. Present suggestion implies that phase IV-HP should have a Ag$_2$CrO$_4$ structure. However, as noted before, phase IV transformed to Na$_2$CrO$_4$ structure, not  Ag$_2$CrO$_4$ structure, during optimization above 10 GPa. This seems contradict with our interpretation. However, it should be noted that stable phase is ultimately defined to have lowest free energy, not by structural relationship. Similarly, structural similarity between olivine and phase III would suggest that olivine structure as structure candidate for III-HP, however, this is less likely from point of the enthalpies and volumes. In order to confirm present interpretation, the quasi-harmonic approximation (QHA) calculation which accounts the vibrational entropy contribution is necessary. Also, experimental study such as in-situ X-ray diffraction study under pressure will ultimately solve the issues.

\subsubsection{Phase II to spinel transition and its relationship with nitrides}
The transformation from phase II to spinel structure was observed during the DFT calculation at 24 GPa. Literature survey reveals that structural relation of phase II-like structure and spinel is well known for nitrides \cite{Kroll}. First-principles study revealed that C$_3$N$_4$ could adapt phase II-like structure with low compressibility \cite{Teter}. They showed that when Zn and Si are replaced with C, and O with N in phase II, the structure slightly relaxes, resulting a cubic phase (designated as willemite II, or ``wII'' in the literature). Later, Zerr et al. \cite{Zerr} and Kroll \cite{Kroll} calculated phase wII of Si$_3$N$_4$ as well. Zerr et al. \cite{Zerr} experimentally synthesized spinel phase of Si$_3$N$_4$ above 15 GPa using a laser-heated diamond anvil cell high-pressure device. However, phase wII has not been experimentally synthesized to date. Kroll \cite{Kroll} calculated relative enthalpies of wII and spinel phases for Si$_3$N$_4$, Ge$_3$N$_4$ and C$_3$N$_4$, and proposed a diffusionless transition mechanism from spinel to phase wII, and also calculated activation energy for the transition. Kroll \cite{Kroll} explained this transition mechanism based on the Bain correspondence, which connects the crystal lattice of face-centered cubic (fcc) and body-centered cubic (bcc) by a homogeneous strain. The anion (oxygen or nitrogen) arrangement in phase II (and wII) is nearly bcc \cite{Marumo}, whereas that of spinel phase is fcc \cite{Sasaki}. Therefore, the Bain correspondence relates anion arrangements of two phases, and provides the transition pathway from spinel to phase wII \cite{Kroll}. 

Accordingly, our observed ``transition'' from phase II to spinel during structural optimization is following this proposed transition route backward. It is interesting to note that this transformation brought Zn from a tetrahedral site to an octahedral site, while Si remains in the tetrahedral site, resulting the normal spinel structure, not inverse spinel structure. For the nitrides, this distinction is not relevant. Spinel phase of Zn$_2$SiO$_4$ has not been experimentally synthesized to date, since at pressures higher than stability field of phase V, it breaks down to ZnSiO$_3$ ilmenite plus ZnO (NaCl structure) \cite{Ito}. The situation is just opposite to Si$_3$N$_4$, in which spinel phase can be synthesized \cite{Zerr}, but phase wII cannot be synthesized thus far. Kroll \cite{Kroll} proposed a metastable route to obtain phase wII starting from spinel phase. Similarly, we propose that spinel phase of Zn$_2$SiO$_4$ can metastably produced by compressing phase II at low temperature. The transition from phase II to spinel will be expected at around 9 GPa based on the enthalpy crossover between two phases as shown in Figure 6. Considering close structural relationship between Si$_3$N$_4$ and Zn$_2$SiO$_4$, crystal chemical insights obtained by present study might be useful to further explore stable and metastable phases of Si$_3$N$_4$, Ge$_3$N$_4$ and C$_3$N$_4$.

\section{Conclusions}
Phase relation of Zn$_2$SiO$_4$ has been studies using the DFT calculations up to 25 GPa. Total 11 phases were considered. Phase III with tetrahedral olivine structure exhibited extremely high compressibility, which is due to reduction of volumes of vacant octahedral sites. We confirmed that phases III and IV recovered from high-pressure synthetic experiments are retrograde phases. From the calculated enthalpies and volumes, we suggested that the high-pressure phase of III would have a Na$_2$CrO$_4$ structure, whereas that of phase IV would have a Ag$_2$CrO$_4$ structure. Therefore, the pressure-induced phase transition sequence of Zn$_2$SiO$_4$ will be: phase I (willemite)$\rightarrow$phase II$\rightarrow$Na$_2$CrO$_4$ phase$\rightarrow$Ag$_2$CrO$_4$$\rightarrow$phase V(modified spinel structure)$\rightarrow$ZnSiO$_3$-ilmenite plus ZnO (NaCl structure). This sequence is quite different to those of other M$_2$SiO$_4$ systems, and only common structure is modified spinel (that realized in Mg$_2$SiO$_4$ and Co$_2$SiO$_4$ only). This peculiar behavior is likely due to closed 3$\it{d}$ orbital inherent to Zn. A transition from phase II to spinel structure was observed during structural optimization at 25 GPa, and the transition is essentially identical to known transition for nitrides. Metastable route to synthesize spinel phase was suggested. 
 
\begin{acknowledgments}
This study was supported by the regular operational grant from Okayama University.
\end{acknowledgments}


\bibliography{zn2sio4fp}

\end{document}